\documentclass[12pt]{article}
\usepackage{graphicx}
\usepackage{hyperref}

\newcommand\pubnumber{SNSN-323-63}
\newcommand\pubdate{\today}

\def\institute{Department of Physics\\The Ohio State University\\
191 West Woodruff Ave\\
Columbus, OH 43210, USA}

\def\Title#1{\begin{center} {\Large #1 } \end{center}}
\def\Author#1{\begin{center}{ \sc #1} \end{center}}
\def\Address#1{\begin{center}{ \it #1} \end{center}}

\newcommand\pubblock{\rightline{\begin{tabular}{l} \pubnumber\\
         \pubdate  \end{tabular}}}
\newenvironment{Abstract}{\begin{quotation}  }{\end{quotation}}
\newenvironment{Presented}{\begin{quotation} \begin{center} 
             PRESENTED AT\end{center}\bigskip 
      \begin{center}\begin{large}}{\end{large}\end{center} \end{quotation}}
\def\Acknowledgements{\bigskip  \bigskip \begin{center} \begin{large}
             \bf ACKNOWLEDGEMENTS \end{large}\end{center}}
\def\beq{\begin{equation}}
\def\eeq#1{\label{#1}\end{equation}}
\def\eeqn{\end{equation}}

\def\beqa{\begin{eqnarray}}
\def\eeqa#1{\label{#1}\end{eqnarray}}
\def\eeqan{\end{eqnarray}}

\let\bar=\overbar

\def\Dslash{\not{\hbox{\kern-4pt $D$}}}
\def\dslash{\not{\hbox{\kern-2pt $\del$}}}

\def\msb{{\bar{\ssstyle M \kern -1pt S}}}

\def\ttbar{\ensuremath{\mathrm{t\overline{t}}}}
\def\bbbar{\ensuremath{\mathrm{b\overline{b}}}}
\def\stat{\ensuremath{\mathrm{stat}}}

\newcommand{\ctW}{\ensuremath{c_{\mathrm{tW}}}}

\newcommand{\ctp}{\ensuremath{c_{\mathrm{t\varphi}}}}
\newcommand{\cpQM}{\ensuremath{c^{-}_{\mathrm{\varphi Q}}}}
\newcommand{\ctG}{\ensuremath{c_{\mathrm{tG}}}}

\newcommand{\cpt}{\ensuremath{c_{\mathrm{\varphi t}}}}

\begin{document}
\begin{titlepage}
\pubblock

\vfill
\Title{Using associated top quark production to probe for new physics within the framework of effective field theory}
\vfill
\Author{Brent R. Yates}
\Address{\institute}
\vfill
\begin{Abstract}
Signs of new physics are probed in the context of an Effective Field Theory using events containing one or more top quarks in association with additional leptons. Data consisting of proton-proton collisions at a center-of-mass energy of $\sqrt{s}=$13 TeV was collected at the LHC by the CMS experiment in 2017. We apply a novel technique to parameterize 16 dimension-six EFT operators in terms of the respective Wilson coefficients (WCs). A simultaneous fit is performed to the data in order to extract the two standard deviation confidence intervals (CIs) of the 16 WCs. The Standard Model value of zero is completely contained in most CIs, and is not excluded by a statistically significant amount in any interval.
\end{Abstract}
\vfill
\begin{Presented}
$13^\mathrm{th}$ International Workshop on Top Quark Physics\\
Durham, UK (videoconference), 14--18 September, 2020
\end{Presented}
\vfill
\end{titlepage}
\def\thefootnote{\fnsymbol{footnote}}
\setcounter{footnote}{0}

\section{Introduction}

The Standard Model (SM) of particle physics is one of the most complete and precise models to date, but it only accounts for 5\% of the known universe. The SM currently provides no correct explanation for dark matter and dark energy, the hierarchy problem, and baryon asymmetry, to name a few. The Large Hadron Collider (LHC) located at CERN can currently probe a center-of-mass energy of $\sqrt{s}=$13 TeV. Therefore, the natural question arises: what if new physics beyond the SM occurs at an energy scale above what the LHC can probe directly? The formalism of Effective Field Theory (EFT) allows us to approximate new physics above this scale purely in terms of SM fields. The strength of each new physics operator ($\mathcal{O}$) is controlled by the so called Wilson coefficients (WCs), and are suppressed by powers of the energy scale $\Lambda$. The effective Lagrangian may be written as
\begin{equation}
	\mathcal{L}_{\mathrm{EFT}} = \mathcal{L}_{\mathrm{SM}} + \sum_{d,i} \frac{c_i^{(d)}}{\Lambda^{d-4}}\mathcal{O}_{i}^{(d)},
\end{equation}
where $\mathcal{L}_{\mathrm{SM}}$ is the SM Lagrangian, $c_{i}$ is the $i^{\mathrm{th}}$ WC, and $d$ is the dimension of the operator. It is important to note that all odd numbered dimensions violate lepton and/or baryon number conservation. This analysis focuses on dimension six; higher dimensions are suppressed by additional powers of $\Lambda$, making them unimportant at this level of precision.\\

The analysis described in this proceeding uses a novel technique to examine data collected by the CMS experiment in 2017, corresponding to an integrated luminosity of $41.5\,\mathrm{fb^{-1}}$. It performs a global fit across all processes---including signal and background. We specifically probe EFT effects using multilepton final states. The procedure used helps to constrain the systematic uncertainties, and any correlations rely solely on the data---no assumptions are made. The production channels examined are: $\mathrm{\ttbar l\nu}$, $\mathrm{\ttbar l\overline{l}}$, $\mathrm{t l\overline{l}q}$, and $\mathrm{tHq}$, where $\mathrm{H \to \bbbar}$ is specifically removed. The complete details of this analysis may be found in~\cite{Sirunyan:2020tqm}.\\

\section{Parameterization of the EFT}

The EFT may be parameterized in simulations by splitting the matrix elements ($\mathcal{M}$) into SM and EFT terms
\begin{equation}
	\mathcal{M} = \mathcal{M}_{\mathrm{SM}} + \sum_{j} \frac{c_j}{\Lambda^2} \mathcal{M}_j.
\end{equation}
The cross section is proportional to $\mathcal{M}^2$, and each simulated event may be viewed as a differential piece of the cross section with an event weight $w$. Therefore, we may parameterize these weights using
\begin{equation}
	w_i\left(\frac{\vec{c}}{\Lambda^2}\right) = s_{0i} + \sum_js_{1ij}\frac{c_j}{\Lambda^2} + \sum_js_{2ij}\frac{c_j^2}{\Lambda^4} + \sum_{j,k}s_{3ijk} \frac{c_j}{\Lambda^2} \frac{c_k}{\Lambda^2},
\end{equation}
where the structure constants ($s$) correspond to: the SM term ($s_0$), interference between the SM and EFT ($s_1$), pure EFT terms ($s_2$), and interference between EFT terms ($s_3$). These weights may be summed to produce the predicted event yields as a function of the WCs.\\

Simulations are generated with non-zero WC values at leading order, and extra partons are included when possible to improve our sensitivity. Initial values are chosen to include all relevant phase space and to optimize the statistical power---$\sigma^2_{\stat} = \sum w^2_i(\vec{c})$. The weight of each event accounts for variations in the yield due to EFT effects, and are used to solve for the structure constants in the quadratic parameterization. These quadratic functions are then used to fit to the data.\\

The simulations are made using the dim6TopEFT model~\cite{AguilarSaavedra:2018nen}. %This model uses the Warsaw basis of dimension-six operators, an energy scale of $\Lambda=1$ TeV, the CKM matrix is assumed to be a unit matrix, the u, d, s, c, e, and $\mu$ masses are all set to zero, the unitary gauge is used and all Goldstone bosons are removed, baryon and lepton number violating operators are removed, and lepton universality us assumed. 
Due to limitations in the model, only tree-level simulations are possible. The 16 operators which have the largest impact on the signal processes, and relatively small impact on the $\ttbar$ background, are considered. Only the real components are considered since the imaginary coefficients lead to CP violation, and are well constrained by EDM experiments and $\mathrm{B} \to X_s \gamma$ decays.

\section{Event selection and signal extraction}

The analysis is split into 35 sub-categories, including: lepton ($\ell$) multiplicity, sum of the lepton charges, jet multiplicity, and b-tagged jet multiplicity. A BDT is applied to help separate the prompt leptons from the non-prompt leptons. All final-state observables are an admixture of the processes---the method does not require we separate the states. Each analysis sub-category stores the sum of the quadratic coefficients, and therefore the event yields are fully parameterized by the WCs. Table~\ref{tab:Categories} lists all the categories used.\\

\begin{table}[htbp]
	\caption{Requirements for the different event categories.  Requirements separated by commas indicate a division into subcategories.  The b jet requirement on individual jets varies based on the lepton category, as described in the text.}
	\label{tab:Categories}
	\resizebox{\linewidth}{!}{
		\begin{tabular}{l c c c c}
			\hline
			Selection              & 2$\ell$ss                                  & \multicolumn{2}{c}{$3\ell$} & $\ge$4$\ell$      \\ \hline
			Leptons                & Exactly 2 leptons                             & \multicolumn{2}{c}{Exactly 3 leptons}         & $\ge$4 leptons      \\
			Charge requirements    & $\sum_{\ell} q <0, \sum_{\ell} q > 0$ & $\sum_{\ell} q <0, \sum_{\ell} q > 0$ & - & -      \\
			Jet multiplicity       & 4, 5, 6, $\ge$7 jets                          & \multicolumn{1}{c}{2, 3, 4, $\ge$5 jets}      & 2, 3, 4, $\ge$5 jets &  2, 3, $\ge$4 jets   \\
			Number of b jets & $\ge$2 b jets                           & \multicolumn{1}{c}{1, $\ge$2 b jets}    & 1, $\ge$2 b jets & $\ge$2 b jets \\
			Dilepton mass          & -                                             & $|m_{\ell\ell} - m_{\mathrm{Z}}| > 10$ GeV & $|m_{\ell\ell} - m_{\mathrm{Z}}| \le 10$ GeV & - \\
			\hline
	\end{tabular}}
\end{table}

%\section{Signal extraction}

Each category listed in Table~\ref{tab:Categories} is treated as a Poisson experiment with a probability of obtaining the observed data. A profiled likelihood is used simultaneously fit all categories and is used to extract the 2 standard deviation ($\sigma$) confidence intervals (CIs). Two fitting procedures are used: one where a single WC is fit while the other 15 are treated as unconstrained nuisance parameters, and another where a single WC is fit while the other 15 WCs are fixed to their SM value of zero. The first fitting procedure is the more physical of the two, as there is no reason for new physics to only favor one WC. The second procedure is an extreme scenario where nature has a single WC. The ability to fit this single WC is limited by the lack of knowledge of the other 15.\\

Systematic uncertainties are treated as nuisance parameters in the profiled fit. The most important systematic uncertainties in this analysis are: the misidentified lepton rate estimate, and simulation modeling including  matrix-element parton-shower matching, missing parton uncertainties, and scale uncertainties.

\subparagraph{Misidentified lepton rate estimate}

Contamination from non-prompt leptons entering into the analysis region are to be expected. This is overcome by examining a multijet enriched background region and comparing this to a $\ttbar+\gamma$ enriched background. The limited statistics of the $\ttbar+\gamma$ background is taken into account, and is treated as an additional source of uncertainty.

\subparagraph{Simulation modeling uncertainties}

Uncertainties in the process of matching matrix element simulations to those produced via parton shower models must be accounted for. The leading term in this uncertainty is from matching the extra partons added to the final-state jets. An additional missing parton uncertainty must be applied to any samples which could not be generated with extra partons. This involves comparing leading order EFT effects without extra partons to next-to-leading order SM simulations, and assigning an uncertainty to cover any discrepancies. Finally, the scale uncertainties due to initial- and final-state radiation are taken into account.

\section{Results}

\begin{figure}[!htbp]
	\centering
	\includegraphics[width=0.75\textwidth]{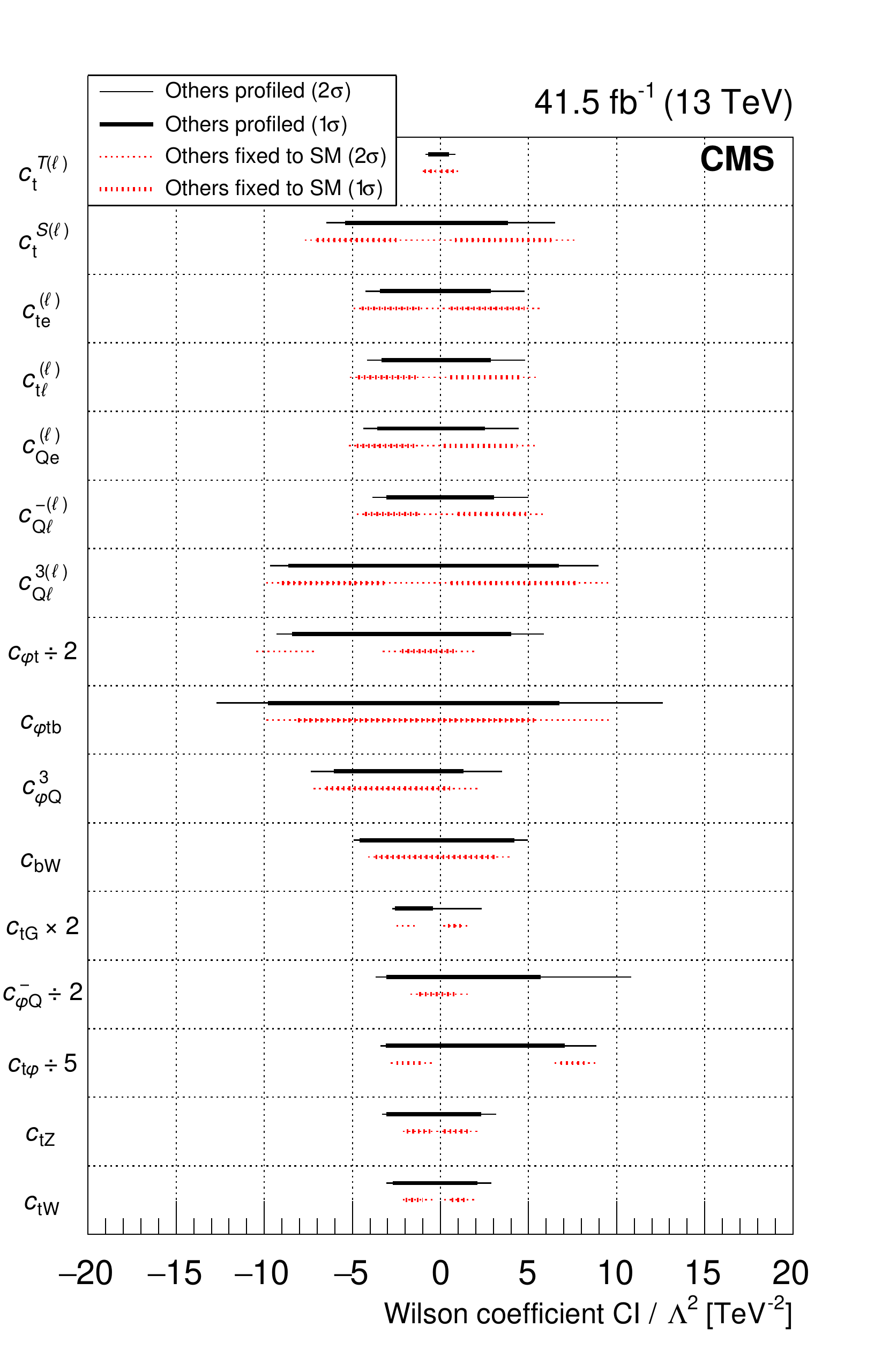}
	\caption{Observed WC 1$\sigma$ (thick line) and 2$\sigma$ (thin line) confidence intervals (CIs). Solid lines correspond to the other WCs profiled, while dashed lines correspond to the other WCs fixed to the SM value of zero. In order to make the figure more readable, the $\cpt$ interval is scaled by $1/2$, the $\ctG$ interval is scaled by 2, the $\cpQM$ interval is scaled by $1/2$, and the $\ctp$ interval is scaled by $1/5$.}
	\label{fig:SummaryPlot}
\end{figure}

\begin{figure}[htbp]
	\centering
	\includegraphics[width=0.85\textwidth]{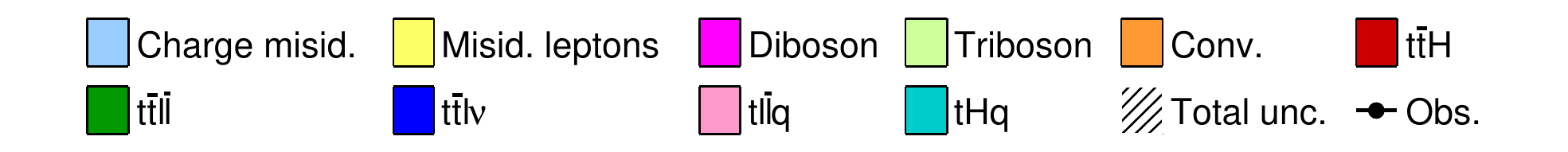} \\
	\includegraphics[width=0.49\textwidth]{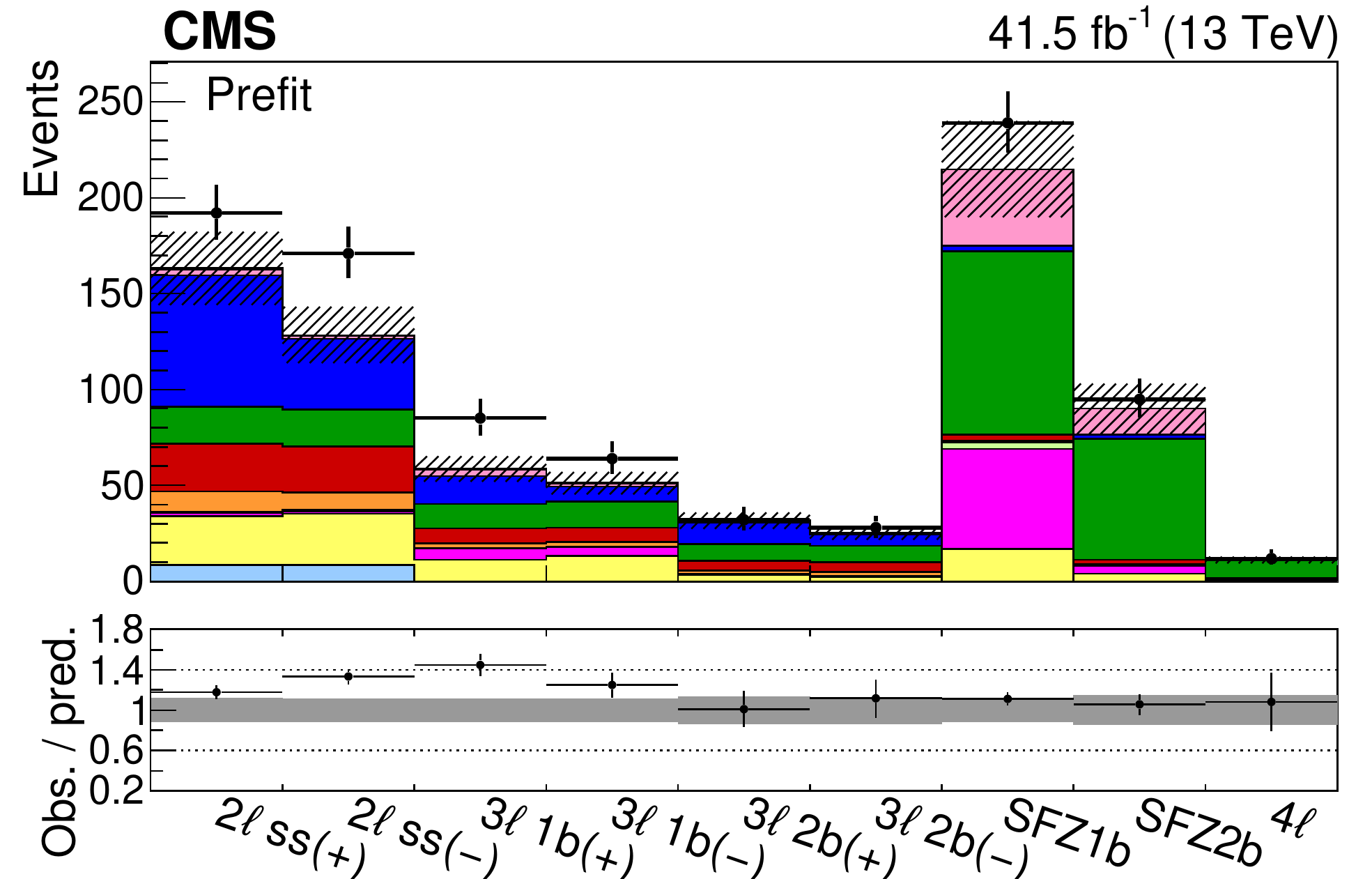}
	\includegraphics[width=0.49\textwidth]{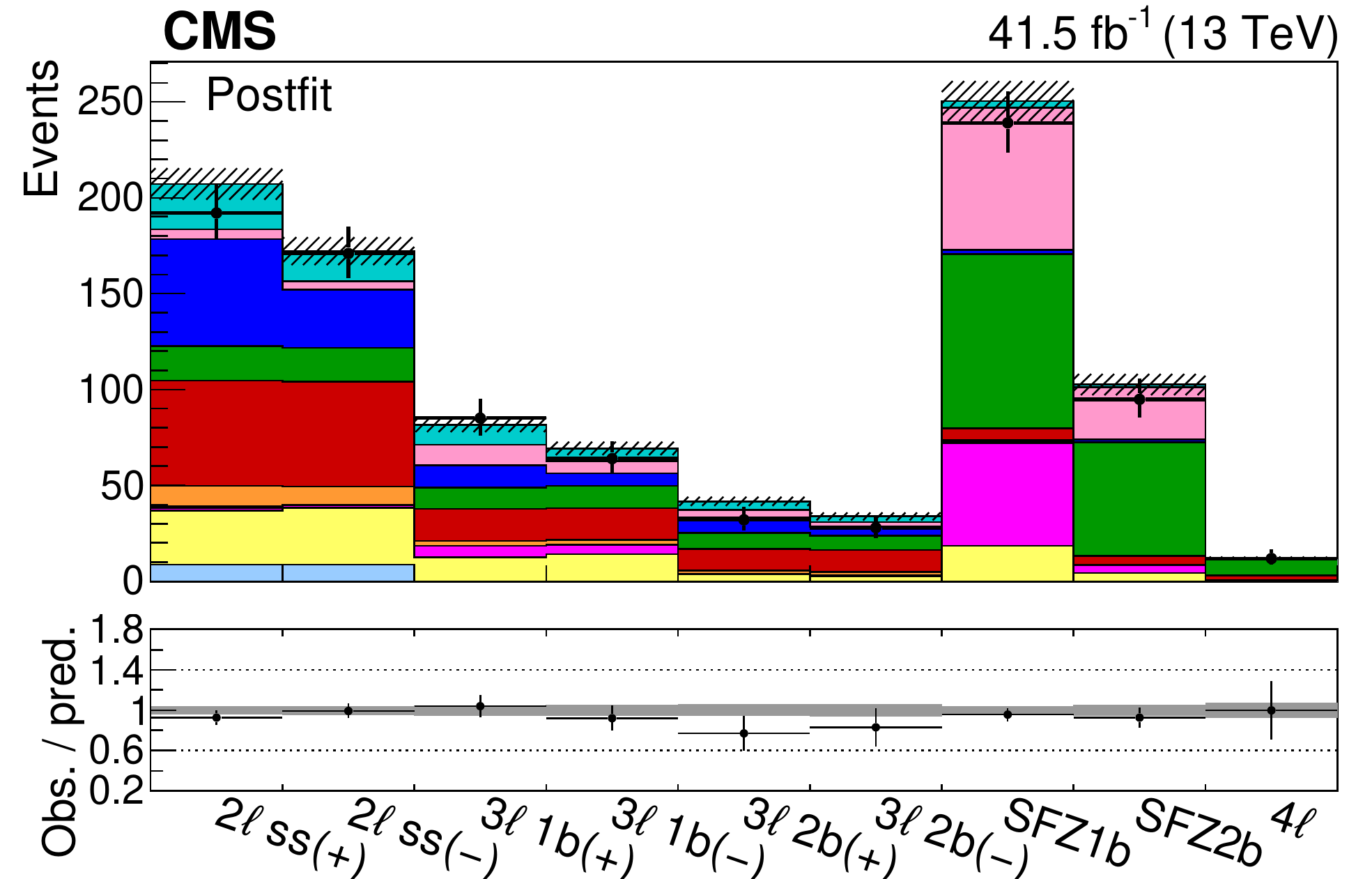}
	\caption{Expected yields prefit (left) and postfit (right). The postfit values of the WCs are obtained from performing the fit over all WCs simultaneously. ``Conv." refers to the photon conversion background, ``Charge misid." is the lepton charge mismeasurement background, and ``Misid. leptons" is the background from misidentified leptons.  The jet multiplicity bins have been combined here, however, the fit is performed using all 35 event categories. The lower panel is the ratio of the observation over the prediction.}
	\label{fig:postfit-yields}
\end{figure}

The 1$\sigma$ and 2$\sigma$ CIs are visualized in Figure~\ref{fig:SummaryPlot}. When the other 15 WCs are fixed to zero $\ctW$, $\ctp$, and $\cpt$ obtain broader disjoint 1$\sigma$ CIs. This is due to the quadratic nature of the parameterization, which broadens the profiled likelihood curves. None of the WCs exclude the SM value of zero by any statistically significant amount. Figure~\ref{fig:postfit-yields} contains the event yields for the SM (left) and the postfit values (right).

\Acknowledgements
We would like to acknowledge the CMS Collaboration for their work in maintaining the CMS experiment and collecting all relevant data for this analysis. We also thank Adam Martin and Jeong Han Kim for their theoretical guidance in configuring and debugging the EFT model used to generate the signal samples in this analysis.

\bibliographystyle{ieeetr}

\bibliography{eprint}

\end{document}